\documentclass[aps,prl,twocolumn,preprintnumbers,groupedaddress]{revtex4}
\usepackage{graphicx}
\usepackage{graphics}

\newcommand{\lsim}{\raisebox{0.6mm}{$\, <$} \hspace{-3.0mm}\raisebox{-1.5mm}{\em $\sim \,$}}

\newcommand{\stau}{\tilde{\tau}}
\newcommand{\mET}{\ooalign{\hfil/\hfil\crcr$E_{\rm T}$}}

\newcommand{\pt}{p_{\rm T}}

\begin{document}

\preprint{UT-09-06, IPMU-09-0026} 

\title{Measuring lifetimes of long-lived charged massive particles stopped in LHC detectors}

\author{Shoji Asai$^1$}
\author{Koichi Hamaguchi$^{1,2}$}
\author{Satoshi Shirai$^1$}
\affiliation{\it {}$^1$ Department of Physics, University of Tokyo, Tokyo 113-0033, Japan\\
{}$^2$ Institute for the Physics and Mathematics of the Universe, 
University of Tokyo, Chiba 277-8568, Japan
}


\begin{abstract}
Long-lived charged massive particles (CHAMPs) appear in various particle physics models beyond the Standard Model.
In this Letter, we discuss the prospects for studying the stopping and decaying events of such long-lived CHAMPs at the LHC detectors, and show that the lifetime measurement (and the study of decay products) is possible with the LHC detectors for a wide range of the lifetime ${\cal O}(0.1)$--${\cal O}(10^{10})$ sec,
by using periods of no $pp$ collision. 
Even a short lifetime of order one second can be measured by (i) identifying the stopping event with the online Event Filter, (ii) immediately making a beam-dump signal which stops the $pp$ collision of the LHC, and at the same time (iii) changing the trigger menu to optimize it
for the detection of  a CHAMP decay in the calorimeter.
Other possibilities are also discussed. 
\end{abstract}

\maketitle
{\bf{\em Introduction.~}}
Long-lived charged massive particles (CHAMPs) appear in various particle physics models beyond the Standard Model (SM)~\cite{Fairbairn:2006gg}. 
Recently there has been growing interests in such a long-lived (charged) particle with a lifetime of ${\cal O}(1000)$ sec, 
since it may solve the lithium problem in big-bang nucleosynthesis (BBN)~\cite{Lithium,Kamimura:2008fx}. 
A well-known example of a CHAMP is a slepton, such as the stau, in supersymmetric (SUSY) models, 
which can be long-lived depending on the SUSY mass spectrum~\cite{Fairbairn:2006gg}.
In this Letter, we discuss the prospects for studying such long-lived CHAMPs with the LHC detectors,
particularly its lifetime measurement.

At the LHC, the CHAMPs are typically generated as a result of cascade decays.
The collider signatures of a long-lived CHAMP very much depend on its lifetime.
(I) If the decay length is short, $c\tau\ll {\cal O}({\rm cm})$, it decays promptly.
(II) For ${\cal O}({\rm cm}) \lsim c\tau \lsim L$, where $L$ is a typical detector size, 
a displaced vertex of the CHAMP decay can be seen inside the detectors. 
There are already some studies of the in-flight decay of CHAMPs, 
and of measurement of CHAMPs' lifetimes, using LHC detectors~\cite{Ambrosanio:2000ik,Asai:2008sk,Ishiwata:2008tp}.
A recent study~\cite{Ishiwata:2008tp} for some SUSY models has shown that a sizable number of 
such long-lived particles can decay inside the detector if its lifetime is shorter than $10^{-(3-5)}$ sec.
(III) If the CHAMP lifetime is even longer, there is no in-flight decay inside the detectors,
and most of the events have massive charged tracks.
The target of this study is this last scenario, case (III).
Hereafter, we call a case (III) CHAMP simply a long-lived CHAMP.

There have also been many studies of long-lived CHAMPs at the LHC~\cite{Fairbairn:2006gg,Hinchliffe:1998ys,Ambrosanio:2000ik,Ellis:2006vu,otherSTAUatLHC}. 
Detailed studies have shown that the CHAMP mass can be measured with an accuracy of 
less than 1\%, with the time-of-flight (TOF) information 
using the muon system~\cite{Ambrosanio:2000ik,Ellis:2006vu}. 
Furthermore, masses and other properties (such as spin and flavor structure) of the CHAMP,
as well as other new particles, can be studied
by using kinematical information~\cite{Hinchliffe:1998ys,Ambrosanio:2000ik,Ellis:2006vu,otherSTAUatLHC}.

Once such a long-lived CHAMP is discovered at the LHC, 
the next important physics target will be the discovery of the CHAMP decay, 
and its lifetime measurement~\footnote{A CHAMP cannot be completely stable 
since its abundance is severely constrained by searches in sea water 
--- even an inflation with a reheating temperature as low as 1 MeV 
cannot dilute the CHAMP abundance enough to avoid the experimental bounds, 
unless its mass is much heavier than 1 TeV~\cite{Kudo:2001ie}.}.
The importance of the lifetime measurement is twofold; 
(A) It is of great importance to cosmology.
The lifetime of the CHAMP and its abundance when it decays in the early universe 
are subject to severe cosmological constraints,
in particular those from BBN~\cite{bounds1,CBBN,Kamimura:2008fx}.
In addition, as mentioned above,
a CHAMP with a lifetime of ${\cal O}(1000)$ sec may solve the lithium problem~\cite{Lithium,Kamimura:2008fx}.
(B) The lifetime measurement will give us precious information on the underlying theory of particle physics.
A popular example of a CHAMP is the stau, in models for which the stau is the next-to-lightest SUSY 
particle (NLSP) and the gravitino is the lightest SUSY particle (LSP). 
In this case, the lifetime $\tau_X$ of the NLSP directly tells us the SUSY breaking scale $\Lambda$, 
or equivalently the gravitino mass $m_{3/2}\approx \Lambda^2/ M_{\mathrm P}$, through the formula
$\tau_X = 6\times 10^4~{\rm sec}~(m_X/100~{\rm GeV})^{-5} (m_{3/2}/1~{\rm GeV})^2 (1-m_{3/2}^2/m_X^2)^{-4}$,
where $M_{\mathrm P} = 2.4\times 10^{18}~{\rm GeV}$ is the reduced Planck scale and $m_X$ is the NLSP mass.
Furthermore, along with an independent kinematical reconstruction of $m_{3/2}$, 
it can lead to a new microscopic determination of the Planck scale, which will 
be an unequivocal test of supergravity~\cite{Buchmuller:2004rq}. 
Even if the underlying theory is not SUSY 
(e.g. the UED model of \cite{Feng:2003nr}),
the lifetime measurement and the study of the decay products 
will be a crucial step forward revealing the underlying theory.

The lifetime measurement of a long-lived CHAMP is, however, very challenging 
because most of the produced CHAMPs typically have large velocities and
escape from the detectors.
Possibilities for stopping and collecting long-lived CHAMPs by 
placing additional material outside the detectors have been discussed~\cite{stop_stau}, 
but they require a new stopper-detector.
(cf. \cite{ILC} for ILC studies.)

The ATLAS detector has massive hadronic calorimeters (HCAL);
in the barrel ($|\eta|\!<\!1.5$) consisting of iron with thickness 1440 mm,
and in the end-cap ($|\eta|\! =\! 1.5-2.5$) consisting of copper with thickness 1400 mm.
Some fraction of the CHAMPs, which are emitted with small velocity $\beta$, lose their kinetic energy 
by ionization 
and will stop inside the HCAL (cf.~\cite{Arvanitaki:2005nq}).
However, the decay of the stopped CHAMPs are out-of-time with beam collisions,
and their decay products do not originate from the beam interaction point, 
and hence the background (BG) rejection will be difficult during $pp$ collisions. 
Actually, it is difficult to trigger the decays of stopped CHAMPs during $pp$ collisions, 
since the triggers of the detectors are optimized for the physics of $pp$ collisions.

We discuss the possibility of discovering and studying the decays of stopped CHAMPs inside 
the ATLAS detector during periods of no $pp$ collisions.
It suffers from much less BG than with $pp$ collisions, 
and we can also optimize the trigger menu for the CHAMP decay.
We show that the lifetime measurement (and the study of decay products) is indeed 
possible for a very wide range of the lifetime ${\cal O}(0.1)$--${\cal O}(10^{10})$ sec. 
One may wonder whether such a study is difficult for a short lifetime, 
since the stopped CHAMPs decay before the beam is switched off. 
However, we show that even a short lifetime of ${\cal O}(1)$ sec can be measured by 
(i) identifying the stopping event by the online Event Filter, 
(ii) immediately making a beam-dump signal to stop collisions,
and at the same time (iii) changing the trigger menu to optimize it
for the detection of  a CHAMP decay in the calorimeter.
We also briefly discuss other possibilities.

Although the basic idea of studying stopped particle decays is model-independent,
for concreteness we assume a SUSY benchmark point 
SPS7 \cite{Allanach:2002nj}, which contains the stau NLSP ($\tilde{\tau}$) as a long-lived CHAMP, with 
the gravitino mass (and hence the stau lifetime) taken as a free parameter. 
In the following analysis, the event selection is optimized for
this specific model, but our basic idea of using stopped
particle decays may also be applicable to other cases
by appropriately changing the selection
cuts, once typical signatures of the new physics are known.
The mass spectra are calculated
by ISAJET 7.78 \cite{ISAJET} and we use Herwig 6.5 \cite{HERWIG6510} and
AcerDET-1.0 \cite{RichterWas:2002ch} to simulate the ATLAS detector.
Precise parameters are used  to emulate the online Event Filter performance~\cite{ATLAS}.
Some of the relevant SUSY particle masses in this model are as follows; 
$(m_{\tilde{\tau}_1},m_{\tilde{g}},m_{\tilde{u}_r})  \!=\! (124,948,871)$\,GeV.
The cross section for SUSY events is approximately 3.5 pb.


{\bf{\em Selection of stopping events.~}}
There are three stages in the ATLAS trigger system.
The event is completely reconstructed in the final
stage of the trigger system, the so called ``Event Filter'' (EF), 
and as such, selection power is equivalent to that of offline analysis.
The following  selections are applied at EF to 
select the events in which the CHAMP stops in the calorimeter.
In each SUSY event, two staus (CHAMPs) are generated as a result of the SUSY cascade decay.

{\bf 1.}
The missing transverse energy $\mET\!\!>\!\! 100$~GeV~\footnote{In 
this model point, the missing transverse energy is mainly caused by the $\tau$ decay.
A slow $\beta$ trigger for the $\stau$ may also be useful, but we do not discuss it here.}. 

{\bf 2.}
The transverse momentum $\pt$ should be larger than 100~GeV for at least one jet.

{\bf 3.}
The number of jets with $\pt \!>\!$ 50~GeV is larger than or equal to 3.
These high-$\pt$ jets are emitted from the cascade decay process.
The first three selections are standard for SUSY studies and suppress the SM BG 
processes significantly. 
The expected BG rate after the first three selections is about 0.02~Hz for 
standard run with a luminosity of $10^{33}{\rm cm}^{-2}{\rm s}^{-1}$.

\begin{figure}
\abovecaptionskip=15pt
\belowcaptionskip=-15pt
\includegraphics[width=6cm]{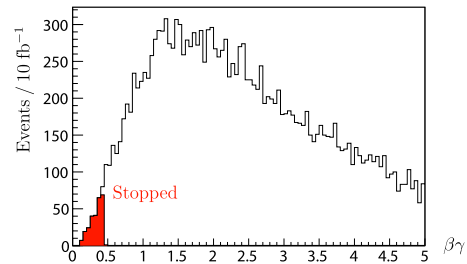}%
\vspace{-.5cm}
\caption{$\beta\gamma$ distribution of the staus. The red histogram shows the stopping stau. Selections 1-4 are applied 
and the numbers are normalized to the integrated luminosity of 10 fb$^{-1}$.
}
\label{fig:st}

\end{figure}

{\bf 4.}
The event should contain one isolated track
whose $\pt$ is larger than 0.1$\times m_{\rm CHAMP}$. 
This track is the candidate for the long-lived CHAMP.
As mentioned above, the mass of the CHAMP can be measured with the TOF information 
from the muon detector.
The $\beta\gamma$ distribution of the staus after the selections 1-4 is shown in Fig.~\ref{fig:st},
for the signal from an integrated luminosity of 10 fb$^{-1}$. 
The red region shows where the CHAMPs stop in the HCAL. 

{\bf 5.}
There should be no track ($\pt \!\!>\!\! 1~{\rm GeV}$) within $\Delta R \!<\!$ 0.3 around 
the candidate track.
This isolation condition is required in order to reduce $\pi^{\pm}$ tracks in jets, 
and a high rejection power, ${\cal O}(10^{3})$, is obtained for high-$\pt$ $\pi^{\pm}$'s. 
Since the SM BG processes are already reduced significantly 
with the first three selections, the BG contribution is expected to be small.

{\bf 6.}
The selected track is extrapolated to the calorimeter system and
the energy deposited on the associated calorimeters is required 
to be smaller than 20\% of the momentum.
This selection rejects  BG tracks from hadrons and electrons.
Since only CHAMPs with small $\beta\gamma$ ($\lsim 0.45$) can stop in the 
calorimeter (see Fig.~\ref{fig:st}), the kinetic energy of the CHAMPs is limited. 

{\bf 7.}
The selected track is also extrapolated to the muon system, and 
no associated track should exist there. 
This selection also reduces the BG from muons.

The events with CHAMPs stopping in the calorimeters are selected with the above selections at EF. 
This online EF selection is crucial for the case of short CHAMP lifetime. 
When a stopping event is detected at EF,
the signal can make a beam-dump command to immediately stop collisions~\footnote{The beam dumping itself is quite a normal operation, done at the end of each run.},
and to switch the trigger menu of the ATLAS detector to 
the dedicated menu for the CHAMP decay.
Since it takes about 1 sec to reconstruct  events in EF and also about 1 sec to change the trigger menu, 
there is a dead time of ${\cal O}(1)$ sec from these procedures.

The trigger rate of the stopping CHAMPs in this model is about 2.3 per day for a luminosity of $10^{33}{\rm cm^{-2}s^{-1}}$ 
(tagging efficiency including stopping probability is about 0.6\%).
This rate is reasonable for the stable operation of the LHC.
The expected fake rate is less than 1/day.

After the beam is dumped, 
we change the trigger menu to detect the tau emitted from the decay of the stau.
The energy of the tau is half that of the NLSP mass and some energy is lost due to $\nu_{\tau}$.  
We select the events in which only one isolated cluster ($>$ 10GeV) exists in the HCAL.
In order to reduce the clusters arising from cosmic rays, we require no track/hit in the muon system.
The BG from radio-active materials and slow neutrons have a small energy of less 
than 1~GeV, and hence we can separate the stau decay event easily from these.
It is important to note that only hadronic decays of the tau are selected because of the following two reasons.
(1) When the tau decays into an electron, the range of the electron is small in the HCAL 
and most of the energy will be deposited in the absorbers (it becomes invisible). 
(2) When the tau decays into a muon, only a small part of the energy will be deposited in the calorimeter.
Also, we have already required no hits on the muon system to reduce the BG from cosmic rays and beam halo.    


{\bf{\em Measurement of the CHAMP Lifetime.~}}
Now let us summarize our setup to measure the CHAMP lifetime.
(i)
After passing EF, the beam is dumped and the trigger menu is switched.
During this process, there is dead time $T_D ={\cal O} (1)$ sec as discussed in the previous section.
(ii)
Then the detector waits for the stau's decay and measures the decay time.
The waiting time $T_W$ would be ${\cal O} (1)$ hour.
(iii)
In addition to the beam running period, the winter shutdown period is also available to detect the 
stau decay~\footnote{There is also a couple of hours break (no $pp$ collision) between each run,
which can be used in a similar way as the winter shutdown discussed in the text.
We do not discuss it here for simplicity.}.
Here, we define $T_R$ and $T_S$ as the lengths of the running and shutdown periods respectively.

Depending on the lifetime of CHAMP, we have different ways to measure it.

{\bf Case I:} 
$\tau_{\tilde{\tau}} \lsim T_W$.
In this case, the decay time can be measured for each event.
We can statistically estimate the value of $\tau_{\tilde{\tau}}$ from the data of the decay time during the waiting time.
Even if  $\tau_{\tilde{\tau}}\ll T_D$, we can estimate the upper bound on $\tau_{\tilde{\tau}}$
\footnote{For $\tau_{\tilde{\tau}} \ll T_D$, the empty-bunch method 
may be useful. See discussion in the next section.}.

{\bf Case II:} 
$T_W \ll \tau_{\tilde{\tau}}\lsim T_S$.
We can measure $\tau_{\tilde{\tau}}$ by using the period of the winter shutdown.
After the LHC running period, 
some amounts of the staus are stored in the calorimeter
and a part of the staus decay in the period of the winter shutdown.
By measuring the decay time during this period, we can estimate $\tau_{\tilde{\tau}}$.

{\bf Case III:}
$T_S \ll  \tau_{\tilde{\tau}}$.
In this case, we can estimate $\tau_{\tilde{\tau}}  \approx\kappa N_S T_S/N_D$, where
$N_D$ is the number of stau decays during the winter shutdown and
$N_S$ the number of all events (including the events which do not pass through the cuts) in which the stau stops in the calorimeter,
and $\kappa$ is the detection efficiency for stau decay.
If we assume that $N_S \approx N_C$, where $N_C$ is the number of events passing the Event Filter,
we can estimate the  $ \tau_{\tilde{\tau}}$.

In Table \ref{tb:error}, we show the expected statistical errors from one year's data.
Here we assume that $T_D = 1$ sec, $T_W = 0.5$ hour, $T_R = 200$ day and $T_S = 100$ day,
and that we can detect all the stau's decay if the tau from $\tilde{\tau}\to \tau+\tilde{G}_{3/2}$
decays hadronically.
The discovery of CHAMP decay and the lifetime measurement is possible for a wide range of 
lifetime ${\cal O}(0.1)-{\cal O}(10^{10})$ sec.

\begin{table} 
\abovecaptionskip=0pt
\belowcaptionskip=0pt
\caption{Expected statistical errors for each lifetime. 
$\left<N_{D} \right>$ is the expected number of staus' decays in the corresponding period.
For 100 fb$^{-1}$ and $\tau_X \simeq {\cal O}(1)$ sec,
the empty-bunch method will be useful. (See discussion below.)}
\label{tb:error}
\begin{tabular}{|c|c|c|c|c|}
\hline
 &\multicolumn{2}{|c|}{10 fb$^{-1}$} &\multicolumn{2}{|c|}{100 fb$^{-1}$}\\ \hline
lifetime &  $\left<N_D\right>$ & $\sigma$ & $\left<N_D\right>$ & $\sigma$\\ \hline
0.1 sec &0.008& $\pm 0.1$ sec &-& - \\ \hline
0.2 sec &1.2& $\pm 0.15$ sec &-&- \\ \hline
0.5 sec &23& $\pm 0.1$ sec &-&- \\ \hline
1 sec &64& $\pm 0.1$ sec &-&- \\ \hline
10 sec &156& $\pm 0.9$ sec &-&- \\ \hline
100 sec &171& $\pm 9$ sec &-&- \\ \hline
1000 sec &144& $^{+ 230}_{-170}$ sec &-&- \\ \hline
10 day &26& $\pm 2.2$ day &262& $\pm 0.7$ day \\ \hline
100 day &143& $^{+49}_{-25}$ day &1430& $^{+20}_{-13}$ day \\ \hline
10 year  &14& $^{+7}_{-3}$ year &138& $^{+1.6}_{-1.2}$ year \\ \hline
50 year &2.8& $^{+110}_{-21}$ year &28& $^{+21}_{-12}$ year \\ \hline
300 year &0.5& $-$  &5& $^{+224}_{-88}$ year \\ \hline
\end{tabular}
\vspace{-0.5cm}
\end{table}

{\bf{\em Discussion.~}}
Although we have taken a SUSY model with the stau NLSP and gravitino LSP as a concrete example, 
the basic idea of our proposal can apply to any models which contain long-lived CHAMPs that are accessible at the LHC. 
Another example in SUSY is the case of an axino LSP, in which the lifetime measurement will probe the Peccei-Quinn scale~\cite{Brandenburg:2005he}.
The right-handed sneutrino LSP scenario can also have a long-lived CHAMP signal~\cite{RHsN}.
The present analysis may also apply to the case of a long-lived colored particle, such as the long-lived gluino (cf. \cite{Arvanitaki:2005nq}), but this case is non-trivial because its behavior in the calorimeters is different from the case of the stau, in particular the track reconstruction may be difficult for charge-flipping R-hadrons.

The study of decay modes, including the particle identifications and the energy measurement, is also very important to reveal the underlying theory. (cf.~\cite{Buchmuller:2004rq,Brandenburg:2005he,RHsN,Hamaguchi:2004ne,decaymodesimplication}.) For the cases that stopped CHAMPs decay mainly into muon and/or electron, the detection will be difficult compared to the case of hadronic decay.
Detailed studies of decay products, including more general cases of CHAMPs (and long-lived colored particles), are beyond the scope of this Letter.

Finally, let us briefly discuss other possibilities alternative to the beam-dump method.

{\bf 1.}
There are empty bunches (about 10 \% of total crossing bunches) in the LHC beam,
and no $pp$ collisions occur during these periods~\cite{C_Hill}.
Both trigger menus for the $pp$ collision and the CHAMPs decay would be working in parallel 
with the bunch information. 
If the CHAMP decays during the empty bunch periods, we can observe it.
The observed number of events is suppressed by a factor of 10, which is serious for 
the precise lifetime measurement.
Furthermore, there is an ambiguity in relating the observed CHAMP decays 
and the produced SUSY events,
especially for a longer lifetime case.
When the gluino mass becomes much lighter and 
the production rate of the stopping CHAMPs  increases significantly,
this empty-bunch method becomes useful.

{\bf 2.}
Changing the beam-orbit with the EF signal;
This method is a modified version of the beam-dump method.
If we can change the orbit of the proton beam quickly and safely (this is an important point),
the EF signal would change the proton beam-orbit  to forbid $pp$ collisions at the ATLAS point 
when the stopping CHAMP is detected.
Such a method would not require a beam-dump and collisions would continue
at other detectors while being suspended at ATLAS~\cite{AJkenkyuukai}.
After the CHAMP's decay is observed in the ATLAS detector, 
collisions can be restarted at ATLAS. 
This method is effective and much more luminosity can be accumulated.
Methods for changing the proton orbit quickly and safely should be developed.
Also we would need a careful study of the beam halo BG for the CHAMP decay.

\begin{acknowledgments}
{\it Acknowledgements}
We thank M.~M.~Hashimoto, C.~Hill, T.~Kawamoto, S.~Komamiya, and H.~Sakamoto for helpful discussions.
SS is supported in part by JSPS Research Fellowships for Young
Scientists. This work is supported by World Premier International
Research Center Initiative (WPI Initiative), MEXT, Japan.

\end{acknowledgments}


\end{document}